\documentclass[twocolumn,aps,prl,showpacs,amsmath,superscriptaddress]{revtex4-1}
\usepackage{amsfonts}
\usepackage{amssymb}
\usepackage{mathrsfs}
\usepackage{graphicx}
\usepackage{float}
\usepackage{xcolor}

\begin{document}

\title{Anderson localization and topological phase transitions in non-Hermitian Aubry-Andr\'{e}-Harper models with p-wave pairing}

\author{Xiaoming Cai}
\address{State Key Laboratory of Magnetic Resonance and Atomic and Molecular Physics, Wuhan Institute of Physics and Mathematics, APM, Chinese Academy of Sciences, Wuhan 430071, China}
\date{\today}

\begin{abstract}
We study non-Hermitian Aubry-Andr\'{e}-Harper models with p-wave pairing, where the non-Hermiticity is introduced by on-site complex quasiperiodic potentials.
By analysing the $\mathcal{PT}$ symmetry breaking, winding numbers of energy spectra, localization and fractal dimensions of states, and fate of Majorana fermions, a complete phase diagram on Anderson localization and topological phase transitions is obtained.
In particular, the non-Hermitian topological nature of Anderson localization phase transitions from extended to critical and then to localized phases is identified, using both analytical and numerical methods.
In the critical phase the complex spectrum is topological nontrivial with a fractional winding number.
In the localized phase the analytical localization length of states can apply to the Hermitian case, which is absent so far.
Both the non-Hermiticity and disorder are detrimental to Majorana fermions.
\end{abstract}
\maketitle

\section{I. Introduction}

Anderson localization (AL) has been one of the most important topics in condensed matter physics \cite{Anderson1958,Abrahams2010}.
In one dimension, an infinitesimal uncorrelated disorder localizes all single-particle states, whereas AL phase transitions can occur at finite strengths in quasiperiodic systems, such as the Aubry-Andr\'{e}-Harper (AAH) model \cite{Aubry1980}.
Recently, given the ability to engineer non-Hermitian Hamiltonians \cite{Peng2016,Xu2016,Weimann2017,Pan2018,Zhou2018}, the interplay between non-Hermiticity and disorder has attracted a great deal of attention, as the non-Hermiticity brings new perspectives on the AL \cite{Hamazaki2020,Okuma2020a,Zhai2020,Claes2020,Hamazaki2019, Wang2019}. 
%
%
In the presence of disorders, non-Hermitian systems exhibit exotic localization phenomena, such as the non-Hermitian skin effect \cite{Leykam2017,Yao2018,Kunst2018,Yokomizo2019,Okuma2020} induced finite-strength localization-delocalization transition \cite{Hatano1996,Silvestrov1999,Longhi2015} and purely imaginary disorder induced AL \cite{Freilikher1994,Asatryan1996,Basiri2014}.
Besides, non-Hermitian quasiperiodic systems, such as various extensions of the AAH model, have also been intensively studied very recently.
The interplay between skin effect and quasiperiodicity leads to asymmetrical AL, and boundary-dependent topologies and self-dualities \cite{Cai2020,Liu2020a,Liu2020b}.
Complex quasiperiodic potentials result in $\mathcal{PT}$ symmetry breaking, topological phase transitions, mobility edges, modified ALs and topological Anderson insulators \cite{Yuce2014,Liu2020c,Longhi2020a,Luo2019,Liang2014,Harter2016, Rivolta2017,Zeng2020a,Longhi2019a,Zhang2020b,Longhi2019b, Zeng2017,Jazaeri2001,Liu2020d,Liu2020e}.
In particular, S. Longhi showed recently that AL phase transitions in non-Hermitian AAH models are of topological nature and are characterized by winding numbers of energy spectra \cite{Longhi2019}.
Is there any other type of system where the AL phase transition is of the non-Hermitian topological nature?
Kitaev chain, a prototype model describing one dimensional (1D) topological superconductors, has attracted a lot of attention, since unpaired Majorana fermions (MFs) are predicted when the system is in the topological phase \cite{Hassan2010,Qi2011,Kitaev2001}.
Due to potential applications in error-free topological quantum computation, many experiments tried to realize the 1D topological superconductor and search for the trace of MFs \cite{Mourik2012,Das2012,Finck2013,Badiane2011,Ioselevich2011}.
Theoretically, various aspects of the Kitaev chain and its extensions have been explored \cite{Wu2012,Zazunov2011,Ueda2014,Cao2012,Liu2011,Lu2014,Katsura2015}.
As to the topological quantum computation, the robustness of MFs against perturbations like disorders is an extremely important issue \cite{Potter2010,DeGottardi2013,Menke2017}.
Previous studies showed that Hermitian quasiperiodic potentials drive the system from the metal phase into a critical phase, and later into the Anderson insulator phase, accompanied by a topological phase transition characterized by the disappearance of unpaired MFs \cite{Cai2013,Wang2016a}.
Subjected to non-Hermitian perturbations, extended Kitaev chains were also discussed recently, with a focus on the $\mathcal{PT}$ symmetry breaking and fate of MFs \cite{Wang2015,Kawabata2018,Klett2017,Li2018,Okuma2019,Liu2020f}.
The AL phase transition in non-Hermitian Kitaev chains, especially its non-Hermitian topological nature and effects on MFs, are not clear yet.
In this paper, we study non-Hermitian AAH models with p-wave pairing, where the non-Hermiticity is introduced by on-site complex quasiperiodic potentials.
The aims are to find out how the AL and topological phase transitions are modified by the non-Hermiticity, to identify the non-Hermitian topological nature of these phase transitions, and to determine the fate of MFs against the non-Hermiticity and disorder.
To these ends, we study the $\mathcal{PT}$ symmetry breaking and winding numbers of energy spectra to identify non-Hermitian topological phase transitions.
The AL phase transition from extended to critical and to localized phases will be clarified by the inverse of participation ratio, using of the fractal theory, and analytical calculation of Lyapunov exponent (inverse of the localization length).
Furthermore, we determine the fate of MFs by the presence of Majorana zero-energy mode (MZM), analytical localization length of MFs, and $Z_2$ topological invariant.
Based on these analyses, a complete phase diagram will be presented.
The rest of paper is organized as follows.
In Sect. II we present the non-Hermitian AAH model with p-wave pairing, its symmetries, the method to solve the model, and the complete phase diagram on AL and topological phase transitions.
The phase diagram will be enriched by studies presented in the next sections.
Sect. III is devoted to studying the $\mathcal{PT}$ symmetry breaking, and non-Hermitian topologies of the energy spectrum which are characterized by winding numbers.
Sect. IV discusses the AL phase transition, Lyapunov exponents, and the critical phase.
In Sect. V we will discuss the fate of MFs and the $Z_2$ topological phase transition.
A summary is provided in Sect. VI.

\section{II. Model, symmetries, and phase diagram}

We consider non-Hermitian AAH models with p-wave pairing, which are described by the following Hamiltonian
\begin{equation}
H=\sum_j(-tc^\dagger_jc_{j+1}+\Delta c_jc_{j+1}+\mathrm{h.c.})+\sum_jV_jc^\dagger_jc_j,
\label{Ham}
\end{equation}
where $c^\dagger_j$ is the creation operator of a spinless fermion at lattice site $j$;
$t$ is the hopping amplitude and set as the energy unit ($t=1$);
$\Delta$ is the p-wave pairing amplitude which can be made positive real, and without loss of generality we will restrict ourselves with $0<\Delta<1$.
The on-site complex quasiperiodic potentials
\begin{equation}
V_j=2V\mathrm{cos}(2\pi\beta j+ih),
\end{equation}
with $V$ the strength.
$\beta$ is an irrational number characterizing the quasiperiodicity.
It usually takes the value of the inverse of golden ratio [$\beta=(\sqrt{5}-1)/2$], which in practice is approximated by rational numbers $\beta=F_n/F_{n+1}$ with $F_n$ the $n$th Fibonacci number.
Correspondingly, the number of lattice sites $L=F_{n+1}$, and in numerical calculations we will take $L=987$ with neglectable finite-size effects.
$h$ characterizes the non-Hermiticity of the system, and we will take it positive real.
The model also can be thought of as Kitaev chains subjecting to complex quasiperiodic potentials.
When $h=0$, the Hermitian disordered Kitaev chain is obtained \cite{Cai2013,Wang2016a}, in which topological and AL phase transitions are well studied.
%
%
%
When $\Delta=0$, the model reduces to the non-Hermitian AAH model \cite{Longhi2019,Cai2020}.
It undergoes a non-Hermitian topological phase transition, where the spectrum changes from real to complex with loops, and accompanied by the AL phase transition.
%


%
The model has $\mathcal{PT}$ symmetry $(\mathcal{PT})H(\mathcal{PT})^{-1}=H$, but not $\mathcal{P}$ and $\mathcal{T}$ symmetries separately, where parity (spatial reflection) $\mathcal{P}$ and time reversal $\mathcal{T}$ operators act as $\mathcal{P}c_j\mathcal{P}^{-1}=c_{L+1-j}$, and $\mathcal{T}i\mathcal{T}^{-1}=-i$.
The $\mathcal{PT}$ symmetry guarantees that the spectrum is made of real or complex conjugate pairs ($E,E^*$) of energies.
In addition, the model also has particle-hole symmetry $ (\mathcal{PC})H(\mathcal{PC})^{-1}=-H$, where the charge conjugation operator $\mathcal{C}$ is defined as $\mathcal{C}c_j\mathcal{C}^{-1}=ic^\dagger_j$, and $\mathcal{C}i\mathcal{C}^{-1}=-i$.
The particle-hole symmetry leads to the presence of ($E,-E^*$) pairs in spectrum.
As a combination of above two symmetries, the model has chiral symmetry $
\mathcal{S}H\mathcal{S}^{-1}=-H$, with operator $S=\mathcal{TC}$.
The chiral symmetry results in the quartet structure ($E,E^*,-E,-E^*$) in spectrum.
According to these symmetries, the model falls into the class BDI in periodic table despite the absence of Hermiticity, and MZMs can exist \cite{Kawabata2019,Zhou2019}.
The Hamiltonian (\ref{Ham}) can be diagonalized by the Bogoliubov-de Gennes (BdG) transformation 
\begin{eqnarray}
\eta^\dagger_n
&=&\sum_{j=1}^L[\phi_{n}(j)\gamma^A_j+i\psi_{n}(j)\gamma^B_j],
\end{eqnarray}
with $n=1,...,L$ the state index.
$\gamma^A_j\equiv c^\dagger_j+c_j$ and $\gamma^B_j\equiv i( c_j-c^\dagger_j)$ are operators of two MFs belonging to one physical site.
They satisfy relations $(\gamma^\kappa_j)^\dagger=\gamma^\kappa_j$ and $\{\gamma^\kappa_j,\gamma^\lambda_k\}=2\delta_{jk}\delta_{\kappa\lambda}$ ($\kappa,\lambda=A,B$) \cite{Kitaev2001}.
Under the BdG transformation, the eigenvalue problem turns into
\begin{eqnarray}
(M-N)(M+N)\phi_n&=&E^2_n\phi_n\notag\\
(M+N)(M-N)\psi_n&=&E^2_n\psi_n,
\label{ENFUN}
\end{eqnarray}
with vectors $\phi_n=[\phi_{n}(1),\phi_{n}(2),...,\phi_{n}(L)]^\mathrm{T}$ and $\psi_n=[\psi_{n}(1),\psi_{n}(2),...,\psi_{n}(L)]^\mathrm{T}$.
The symmetric and antisymmetric tridiagonal matrices $M$ and $N$ are
\begin{equation}
M=\left(\begin{array}{cccc} V_1&-t& &-t\\
-t&V_2&\ddots& \\
 &\ddots&\ddots&-t \\
-t& &-t&V_L
\end{array}
\right), N=\left(\begin{array}{cccc} 0&-\Delta& &\Delta\\
\Delta&0&\ddots& \\
 &\ddots&\ddots&-\Delta \\
-\Delta& &\Delta&0
\end{array}
\right).
\end{equation}
Solving above equations, we obtain the spectrum and all single-particle states ($\phi_n,\psi_n$).
%
%

\begin{figure}[tbp]
\begin{center}
\vspace{0.5cm}
\includegraphics[width=\linewidth, bb=0 0 425 137]{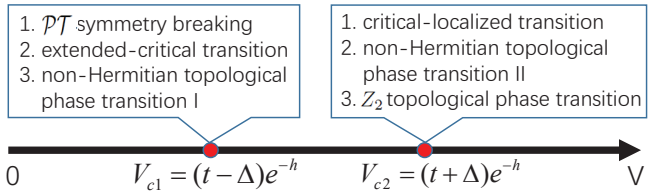}
\caption{Phase diagram of the non-Hermitian AAH model with p-wave pairing.
The non-Hermitian topological phase transition I (II) describes the transition between spectra with and without the loop structure (encircling the origin) in the complex energy plane.}
\label{Fig1}
\end{center}
\end{figure}

%
Properties of the model are concluded into the phase diagram shown in Fig.\ref{Fig1}.
It has two critical points
\begin{equation}
V_{c1(2)}=(t\mp\Delta)e^{-h},
\end{equation}
corresponding to multiple phase transitions.
When $0<V<V_{c1}$, the spectrum of system is real and all bulk single-particle states are extended.
In this extended phase, the system is also topological and has MZMs under the open boundary condition.
Two unpaired MFs are exponentially localized at ends of chain, and the Lyapunov exponent (LE) of them is independent of the disorder strength $V$.
In the intermediate region ($V_{c1}<V<V_{c2}$), the system is in the critical phase, where all bulk states are critical with fractal dimensions.
The $\mathcal{PT}$ symmetry is broken, and the spectrum is complex with loops in the energy plane.
The winding number of spectrum is fractional.
Unpaired MFs still exist but with a disorder-dependent LE.
The transition from real to complex with loops in spectra is named the non-Hermitian topological phase transition I in Fig.\ref{Fig1}.
When $V>V_{c2}$, all states are exponentially localized.
In the localized phase, the $\mathcal{PT}$ symmetry is still broken.
But there is a loop in spectrum encircling the origin of the complex energy plane, which is absent when $V<V_{c2}$.
The changes of the loop structure and winding numbers of energy spectra at $V_{c2}$ define the non-Hermitian topological phase transition II in Fig.\ref{Fig1}.
The loop encircling the origin causes a band inversion and a topological phase transition happens.
The system turns into the topological trivial phase without unpaired MFs when $V>V_{c2}$.
From another perspective, the degree of non-Hermiticity also can drive the above mentioned phase transitions, given a finite $V$.
Both the non-Hermiticity and disorder are detrimental to MFs.
Detailed discussions on AL and topological phase transitions are presented in the following sections.

\section{III. $\mathcal{PT}$ symmetry breaking and winding numbers of spectrum}

\begin{figure}[tbp]
\begin{center}
\includegraphics[scale=0.9, bb=0 0 228 451]{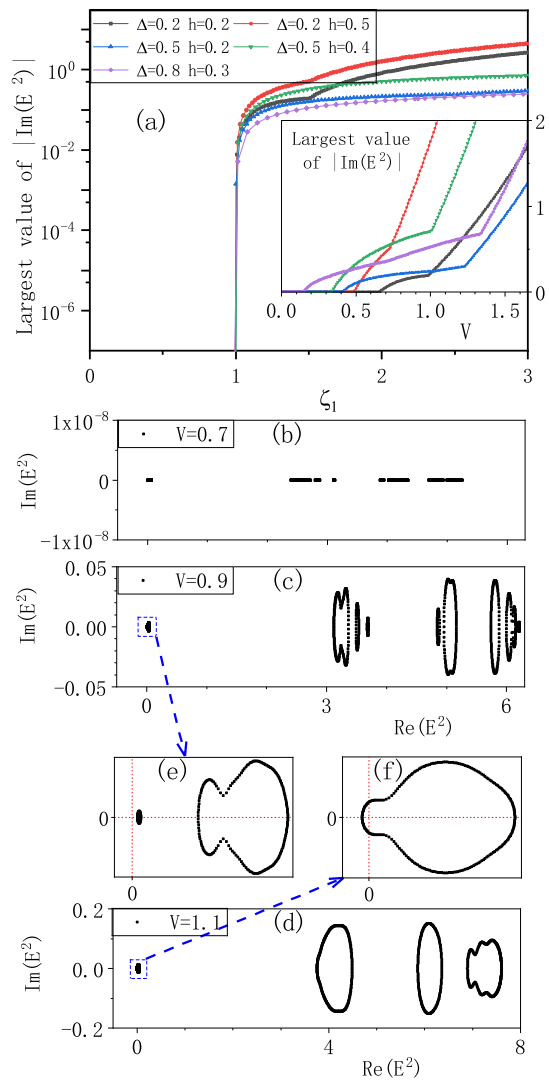}
\caption{$\mathcal{PT}$ symmetry breaking.
(a) Behaviour of the largest values of $|\mathrm{Im}(E^2)|$ vs. $\zeta_1\equiv Ve^h/(t-\Delta)$.
Inset in (a): Corresponding ones vs. $V$.
(b-d) Spectra $E^2$ in the complex energy plane for systems in $\mathcal{PT}$ symmetry (un)broken or different topological phases.
(e,f) Enlargements of (c,d) around the origin, respectively.
$\Delta=h=0.1$ in calculations of (b-f), and corresponding critical points $V_{c1}\simeq0.8144$ and $V_{c2}\simeq0.9953$.}
\label{Fig2}
\end{center}
\vspace{-0.3cm}
\end{figure}

%
%
The complex quasiperiodic potential is an overall balanced gain and loss, and induces the $\mathcal{PT}$ symmetry breaking.
In the inset of Fig.\ref{Fig2} (a) we show maximal values of imaginary parts of all eigenenergies $E^2$ vs. $V$.
As $V$ increases, systems undergo a $\mathcal{PT}$ symmetry breaking phase transition.
After rescaling, in Fig.\ref{Fig2} (a) we present them vs. $\zeta_1\equiv Ve^h/(t-\Delta)$ in the semi-log style.
It clearly shows that phase transition points collapse at $\zeta_1=1$, corresponding to $V=V_{c1}$.
When $V<V_{c1}$, the system is in the $\mathcal{PT}$ symmetry unbroken phase, whereas in the $\mathcal{PT}$ symmetry broken phase when $V>V_{c1}$.

In order to explore more of the spectrum, in Fig.\ref{Fig2} (b-d) we present some exemplary spectra in the complex energy plane.
Here and after we treat Eqs.(\ref{ENFUN}) as the single-particle eigenvalue problem and $E^2$ as the spectrum.
When $V<V_{c1}$ the spectrum is real, and contains bands with each having sub-bands due to the quasiperiodicity of complex potentials.
Whereas when $V>V_{c1}$ the $\mathcal{PT}$ symmetry is broken and the spectrum is complex with loops.
The $\mathcal{PT}$ symmetry is almost completely broken, and only a few eigenenergies have very small imaginary parts due to the presence of loops crossing the real axis.
Further careful analysis shows that when $V>V_{c2}$ there exists a loop encircling the origin of the complex energy plane, which is absent when $V<V_{c2}$ [see Fig.\ref{Fig2} (e,f)].
For the orginal Hamiltonian $H$, whose eigenenergies are $(E,-E)$ paired because of the chiral symmetry, this means that as $V$ increases and crosses the critical point $V_{c2}$ the superconducting gap closes and reopens with a band inversion, which usually induces a topological phase transition (see section V).
Due to the complex nature, the spectrum of non-Hermitian system can have non-trivial topological structures (loops) \cite{Leykam2017,Yao2018,Kunst2018,Yokomizo2019,Okuma2020,Longhi2019,Cai2020}.
To study the topology of spectrum, we introduce an additional dimension by adding a phase $\delta$ in the complex quasiperiodic potential.
Given $V_j=2V\mathrm{cos}(2\pi\beta j+ih+\delta/L)$, winding numbers of energy spectra are defined as \cite{Longhi2019,Cai2020}
\begin{equation}
\nu=\lim\limits_{L\rightarrow\infty}\frac{1}{4\pi i}\int_0^{2\pi}\mathrm{d}\delta\frac{1}{\partial\delta}\mathrm{ln}[\mathrm{det}(H^2-E_B)],
\label{WN}
\end{equation}
which refer to how the complex spectral trajectory $E^2$ encircles the base energy $E_B$ in the complex energy plane, with respect to $\delta$ from $0$ to $2\pi$.
Based on Eqs.(\ref{ENFUN}), in the definition $H^2$ is used instead of $H$.
%
%
And an additional $1/2$ is added in the prefactor.
%
%

\begin{figure}[tbp]
\begin{center}
\hspace{-0.9cm}
\includegraphics[width=\linewidth, bb=0 0 227 164]{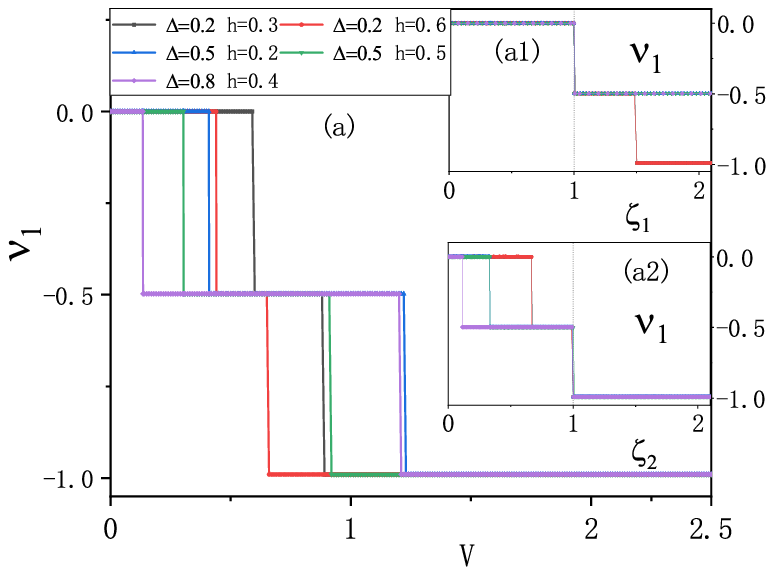}
\includegraphics[width=\linewidth, bb=0 0 227 85]{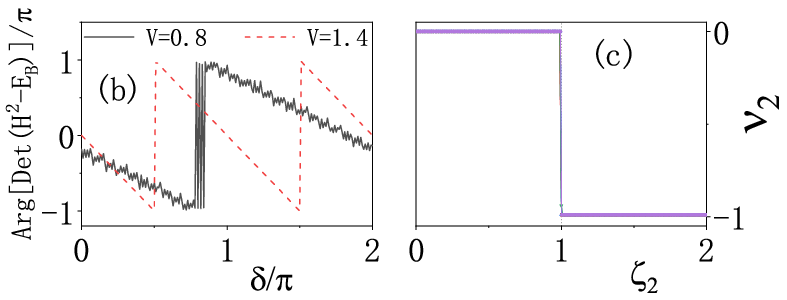}
\caption{Topology of spectrum.
(a) Winding numbers $\nu_1$ vs. $V$, numerically computed using Eq.(\ref{WN}).
Insets (a1-2) in (a): Corresponding $\nu_1$ vs. $\zeta_1$ and $\zeta_2\equiv Ve^h/(t+\Delta)$, respectively.
(b) The phase argument of $\mathrm{det}(H^2-E_B)$ vs. $\delta$ for systems in phases with different $\nu_1$.
$\Delta=0.5$ and $h=0.2$ in calculation of (b), and corresponding critical points $V_{c1}\simeq0.4094$ and $V_{c2}\simeq1.2281$.
(c) Winding numbers $\nu_2$ vs. $\zeta_2$.
The legend of (c) is the same as for (a).}
\label{Fig3}
\end{center}
\end{figure}

Different choices of the base energy $E_B$ give winding numbers characterizing different loop structures.
We concentrate on two cases:
(1) The most non-trivial winding number for any $E_B$, i.e. $\nu_1=\mathrm{sgn}(\nu)\cdot\mathrm{max}(|\nu|),\forall E_B\in\mathbb{C}$, which characterizes the existence of loops in the complex energy plane.
(2) The winding number $\nu_2=\nu|_{E_B=0}$, characterizing loops which encircle the origin.
In Fig.\ref{Fig3} (a) and insets (a1) and (a2) we show $\nu_1$ vs. $V$, $\zeta_1$, and $\zeta_2\equiv V e^{h}/(t+\Delta)$ respectively, numerically computed using Eq.(\ref{WN}).
The winding number
\begin{equation}
\nu_1=\left\{\begin{array}{l}0,\quad\quad\quad 0<V<V_{c1},\\-1/2,\quad V_{c1}<V<V_{c2},\\-1,\quad\quad V>V_{c2},\end{array}\right.
\end{equation}
with two critical points $V_{c1(2)}$, or $\zeta_{1(2)}=1$.
When $V<V_{c1}$, the system is in the $\mathcal{PT}$ symmetry unbroken phase and the spectrum is real with a trivial winding number.
In the middle ($V_{c1}<V<V_{c2}$), the complex spectrum has loops with a fractional winding number.
Fractional winding numbers were reported before in other non-Hermitian systems \cite{Lee2016,Yin2018}.
As $\delta$ increases from $0$ to $2\pi$, the spectral trajectory $E^2$ encircles the base energy once, which is confirmed by the relation between phase argument of $\mathrm{det}(H^2-E_B)$ and $\delta$ [shown in Fig.\ref{Fig3} (b)].
Effectively, the spectrum of $H$ winds half of the circle and has a fractional winding number $-1/2$.
When $V>V_{c2}$, the spectral trajectory $E^2$ encircles twice [see Fig.\ref{Fig3} (b)], and the winding number changes.
On the other hand, we show numerical winding numbers $\nu_2$ vs. $\zeta_2$ in Fig.\ref{Fig3} (c).
The winding number
\begin{equation}
\nu_2=-\theta(\zeta_2-1)=-\theta(V-V_{c2}),
\label{WN2}
\end{equation}
with $\theta(x)$ the step function.
It precisely characterizes the presence of loops encircling the origin.
When $V>V_{c2}$, the spectral trajectory $E^2$ encircles twice around the origin with a non-trivial winding number $-1$.
An analytical proof of Eq.(\ref{WN2}) is given in Appendix A, where the determinant $\mathrm{det}(H^2)$ is obtained from that of the Hermitian AAH model, using an asymmetrical similarity transformation and the Fourier transformation.

\section{IV. Anderson localization and critical phase}

\begin{figure}[tbp]
\begin{center}
\includegraphics[scale=0.9, bb=0 0 228 524]{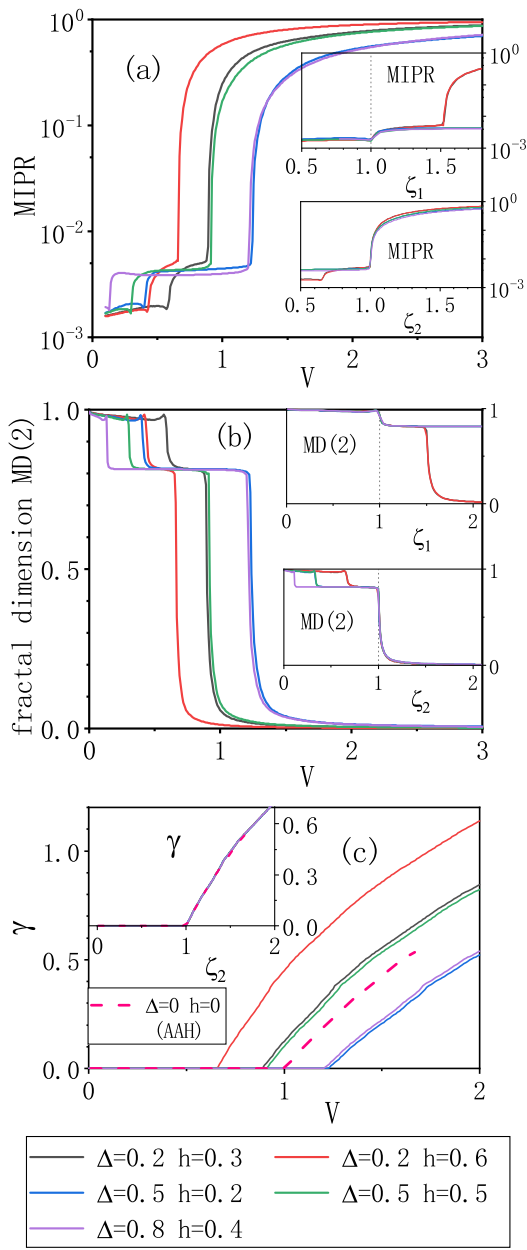}
\caption{Anderson localization and critical phase.
We present mean inverse of the participation ratios (MIPRs), mean multi-fractal dimensions $\mathrm{MD}(2)$, and mean Lyapunov exponents $\gamma$ vs. $V$ in (a), (b), and (c), and corresponding ones vs. $\zeta_{1(2)}$ in insets, respectively.
}
\label{Fig4}
\end{center}
\end{figure}

With an irrational $\beta$, the quasiperiodic potential acts as a disorder and induces the localization of states.
We treat $\phi$ and $\psi$ as the single-particle states, and they have the same localization properties.
The inverse of the participation ratio (IPR) is the most appropriate quantity to characterize the localization of a single-particle state.
We focus on the IPR of $\phi$, which is defined as $P_n=\sum^L_{j=1}|\phi_n(j)|^4$ for a normalized $\phi_n$.
For an extended state the IPR is of the order $1/L$, whereas it approaches to $1$ for a localized state.
In the middle, $P_n\propto L^{-\alpha}$ with $0 <\alpha< 1$, for a critical state which has multi-fractal properties.
In order to characterize the localization of the whole system the mean inverse of the participation ratio (MIPR) $P=\sum_nP_n/L$ is defined.
In Fig.\ref{Fig4} (a) and insets we present semi-log plots of MIPRs vs. $V$, $\zeta_1$, and $\zeta_2$ respectively, for systems under the periodic boundary condition.
There are sudden increases at $\zeta_1=1$ and $\zeta_2=1$, indicating dramatic changes in the localization of states.
The AL phase transition points are the same as the topological ones.
The system is in the extended phase when $V<V_{c1}$, and (M)IPRs$\simeq 1/L$, whereas when $V>V_{c2}$ it is in the localized phase with (M)IPRs$\simeq 1$.
No mobility edge is encountered.
In the intermediate region ($V_{c1}<V<V_{c2}$), MIPRs are significantly larger than $1/L$, but still one order smaller than $1$, which indicates all single-particle states are critical.
To support this statement, we further use the box-counting method to study multi-fractal properties of states.
Dividing a single-particle state into $L/r$ segments with each length $r$, one can define a quantity \cite{Wang2016b}
\begin{equation}
\chi_n(q)=\sum_{m=1}^{L/r}\left[\sum_{j=(m-1)r+1}^{mr}|\phi_n(j)|^2\right]^q.
\end{equation}
Multi-fractal property of the state is characterized by a power-law relation $\chi(q)\sim(r/L)^{\tau(q)}$, where the exponent $\tau(q)$ determines the multi-fractal dimension $D(q)=\tau(q)/(q-1)$ \cite{Dubertrand2014}.
We set $q=2$ as usual.
The power-law relation and a multi-fractal dimension $0<D(2)<1$ are characteristic features of a critical state, while $D(2) = 1$ for an extended state and $D(2)=0$ for a localized state in the thermodynamics limit.
Similarly, we define the mean multi-fractal dimension $\mathrm{MD}(2)$ for the system.
In Fig.\ref{Fig4} (b) and insets we show $\mathrm{MD}(2)$ vs. $V$, $\zeta_{1}$, and $\zeta_{2}$, respectively.
The mean multi-fractal dimensions experience sudden changes at critical points $V_{c1(2)}$.
Considering the finite-size effect, $\mathrm{MD}(2)\simeq 1$ and states are extended when $V<V_{c1}$, whereas when $V>V_{c2}$ states are localized with $\mathrm{MD}(2)\simeq 0$.
In the middle ($V_{c1}<V<V_{c2}$), states are critical.
Now we study the exponential decay of states in the localized phase.
We adopt exponential wave functions $\phi_n(j)=\mathrm{exp}(-\gamma_n|j-j_0|)$ with $j_0$ the localization center and $\gamma_n$ the LE or inverse of localization length.
Extracted by fitting numerical single-particle states with above wave functions, the mean LEs $\gamma=\sum_n\gamma_n/L$ are shown in Fig.\ref{Fig4} (c).
After rescaling, all mean LEs collapse into a single curve with the AL phase transition point $\zeta_2=1$ [see inset in Fig.\ref{Fig4} (c)].
In Fig.\ref{Fig4} (c) and inset we also show the LE for Hermitian AAH model.
Given the LE $\gamma=\mathrm{ln}(V/t)$ for AAH model, we conclude that in the localized phase the LE
\begin{equation}
\gamma=\mathrm{ln}(\zeta_2)=\mathrm{ln}\frac{Ve^h}{t+\Delta},
\label{LEbulk}
\end{equation}
for the non-Hermitian AAH model (\ref{Ham}).
The LEs are energy-independent.
In Appendix B, we analytically prove Eq.(\ref{LEbulk}) by extending to the non-Hermitian realm Thouless's result relating LE to the density of state \cite{Thouless1}.
The LE (\ref{LEbulk}) is also applicable to the Hermitian case ($h=0$), no knowledge about the localization detail has been obtained before \cite{Cai2013}.

\section{V. Fate of Majorana fermions and $Z_2$ topology}

\begin{figure}[tbp]
\begin{center}
\includegraphics[scale=0.9, bb=0 0 228 384]{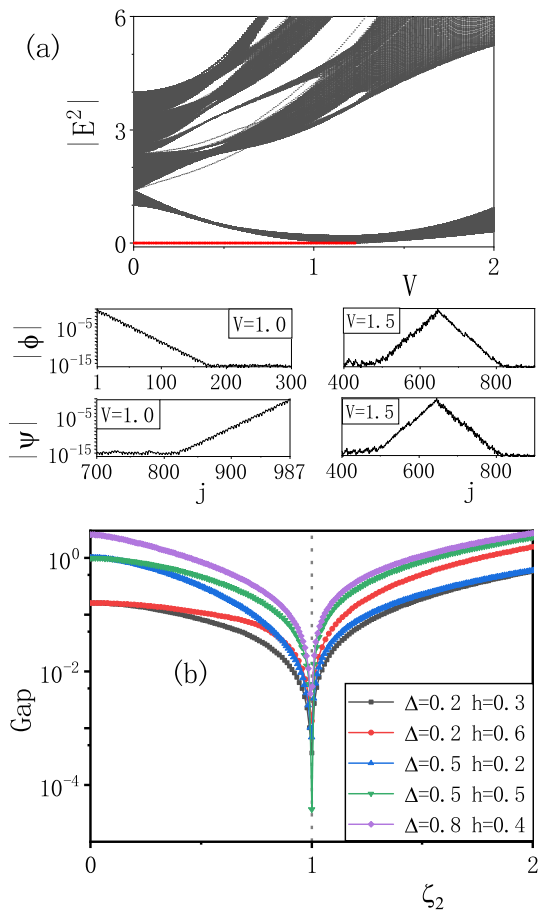}%
\caption{Fate of Majorana fermions.
Top panel in (a): Absolute values of spectra $|E^2|$ vs. $V$ for systems under the open boundary condition.
Bottom panels in (a): Spacial distributions of $\phi$ and $\psi$ for the lowest excitation mode.
$\Delta=0.5$ and $h=0.2$ in calculation of (a), and corresponding critical point $V_{c2}\simeq1.2281$.
(b) Semi-log plot of the absolute bulk energy gap vs. $\zeta_2$.}
\label{Fig5}
\end{center}
\end{figure}

\begin{figure}[tbp]
\begin{center}
\includegraphics[scale=0.9, bb=0 0 228 192]{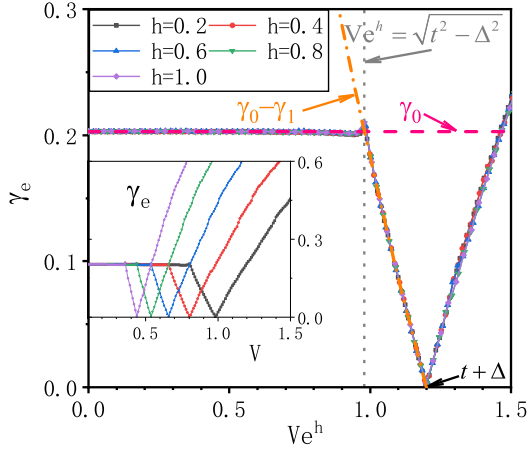}%
\caption{Localization of Majorana fermions.
Lyapunov exponents $\gamma_e$ of the state $\phi$ vs. $Ve^h$ for the lowest excitation mode.
Inset: Corresponding ones vs. $V$.
$\Delta=0.2$ in the numerical calculation.
}
\label{Fig6}
\end{center}
\end{figure}

As mentioned in Sect. II, the model (\ref{Ham}) is in the BDI class and can have MZMs.
In this section we examine the fate of MFs.
In the top panel of Fig.\ref{Fig5} (a) we show typical absolute spectra $|E^2|$ vs. $V$ for systems under the open boundary condition (OBC).
As $V$ increases the absolute bulk energy gap decreases, closes, and reopens again.
Compared with spectra under the periodic boundary condition, an obvious feature is the existence of zero-energy mode in the gapped region before gap-closing point.
This zero-energy mode corresponds to two unpaired MFs, which are exponentially localized at two ends of chain respectively [see bottom panels of Fig.\ref{Fig5} (a)].
The presence of MZM defines the non-trivial topological nature of the system.
In the gap reopened region, no MZM exists and the system is topological trivial.
Two MFs for the lowest excitation mode are exponentially localized in the bulk and overlapped nicely.
Consequently, corresponding quasiparticle is a localized fermion which can not split into two unpaired MFs.
In order to determine the topological phase transition point, in Fig.\ref{Fig5} (b) we show the absolute bulk gap vs. $\zeta_2$ for different systems.
All gap-closing points collapse at $\zeta_2=1$, indicating the topological phase transition point $V_{c2}$.
In addition to the presence of MZMs and gap-closing points, the topological nature is more precisely characterized by the $Z_2$ topological invariant $I=(-1)^\eta$, with $\eta$ the number of MZMs.
Next we use the transfer matrix approach to identify the $Z_2$ topological invariant, LE of MFs, and the topological phase transition.
If the system under OBC hosts a MZM, from Eqs.(\ref{ENFUN}) states $\phi$ and $\psi$ for the zero-energy mode should satisfy $(M+N)\phi=0,(M-N)\psi=0$.
In the transfer matrix form the equation of $\phi$ can be rewritten
as
\begin{equation}
\left(\begin{array}{c}\phi(j+1)\\ \phi(j)\end{array}\right)=T_j\left(\begin{array}{c}\phi(j)\\ \phi(j-1)\end{array}\right),
\end{equation}
with
\begin{equation} T_j=\left(\begin{array}{cc}\frac{2V\mathrm{cos}(2\pi\beta j+ih)}{\Delta+t}&\frac{\Delta-t}{\Delta+t}\\ 1&0\end{array}\right).
\end{equation}
Then the transfer matrix of the whole system $T=\Pi_{j=1}^LT_j$, and we denote two eigenvalues of it by $\lambda_1$ and $\lambda_2$.
Given $0<\Delta<1$, we have $|\mathrm{det}(T)|<1$ and $|\lambda_1\lambda_2|<1$.
If both $|\lambda_1|$ and $|\lambda_2|$ are less than $1$, there is a MZM with normalizable wave functions and the system is in the topological phase.
Otherwise, states are unnormalizable and the system is topological trivial without MZMs.
Supposing $|\lambda_1|<|\lambda_2|$, the $Z_2$ topological invariant $I=-\theta(1-|\lambda_2|)$ and the LE of MF wave functions is defined as $\gamma_e=\lim_{L\rightarrow\infty}\frac{-\mathrm{ln}|\lambda_2|}{L}$.
In order to determine the topological invariant and LE of MFs, we perform a transformation \cite{DeGottardi2013b}
\begin{equation}
T(t,V,\Delta)=\left(\sqrt{\frac{t-\Delta}{t+\Delta}}\right)^LST(t,\frac{V}{\sqrt{t^2-\Delta^2}},0)S^{-1},
\label{TM}
\end{equation}
with $S=\mathrm{diag}(\xi^{1/4},\xi^{-1/4})$ and $\xi=\frac{t-\Delta}{t+\Delta}$.
The matrix $T$ on right side is the transfer matrix of the model with $\Delta=0$ and a renormalized disorder strength.
%
%
According to Ref.\cite{Cai2020}, the LE for transfer matrix on the right side
\begin{equation}
\gamma_1=\mathrm{ln}\frac{Ve^h}{\sqrt{t^2-\Delta^2}},
\end{equation}
when $Ve^h/\sqrt{t^2-\Delta^2}>1$, otherwise, $\gamma_1=0$.
Then following Eq.(\ref{TM}), the LE of MFs
\begin{equation}
\gamma_e=\left\{\begin{array}{l}\gamma_0, \quad\quad\quad\quad\quad\quad Ve^h<\sqrt{t^2-\Delta^2},\\ \gamma_0-\gamma_1,\quad \quad\quad \sqrt{t^2-\Delta^2}<Ve^h<t+\Delta,\\ <0\mathrm{ (unphyical)}, \quad Ve^h>t+\Delta,\end{array}\right.
\label{LE}
\end{equation}
where
\begin{equation}
\gamma_0=\frac{1}{2}\mathrm{ln}\frac{t+\Delta}{t-\Delta},
\end{equation}
is the LE of MFs for the clean Kitaev chain \cite{DeGottardi2013b}.
In Fig.\ref{Fig6} we show numerical LEs for the lowest excitation mode, which agree very well with Eq.(\ref{LE}) when the system is in the topological phase.
From the LE of MFs we obtain the topological invariant through $\lambda_2$
\begin{equation}
I=-\theta(1-\frac{Ve^h}{t+\Delta}),
\end{equation}
which also indicates the topological phase transition point $\zeta_2=1$ or $V_{c2}$.
From above analyses, we also learn that the non-Hermiticity is detrimental to the presence of MFs.
The larger $h$ is, a weaker disorder is needed to drive the system into the topological trivial Anderson localized phase.

\section{VI. Summary}

In summary, we have studied AL and topological phase transitions in the 1D non-Hermitian AAH model with p-wave pairing, where the non-Hermiticity is introduced by on-site complex quasiperiodic potentials.
By analyzing the $\mathcal{PT}$ symmetry breaking, winding numbers of energy spectra, IPRs, fractal dimensions of states, LEs of bulk and edge states, and the existence of MZMs, we determined the complete phase diagram which was presented in Fig.\ref{Fig1}.
Two critical points are identified, corresponding to multiple phase transitions.
As the disorder strength increases and passes through two critical points, the spectrum changes from real to complex with loops in the energy plane, and then to with loops encircling the origin.
Winding numbers of energy spectra change correspondingly.
Extended bulk single-particle states turn into critical states with fractal dimensions, and then into localized states.
We analytically derive the LE of bulk states, which can apply to the Hermitian case where no analytical result has been obtained before.
The exponentially localized MFs become more and more extended, and then disappear into the bulk.
From another point of view, the non-Hermiticity is also detrimental to the presence of MFs.
Increasing the degree of non-Hermiticity also can drive the system into the critical phase, and then into the topological trivial Anderson localized phase, when the disorder strength is finite.
The model can map to a system of two coupled non-Hermitian AAH chains \cite{Wang2016a}, which can be experimentally realized in electric circuits \cite{Cai2020,Jiang2019,Zeng2020b,Longhi2019}.

\section{Acknowledgments}

This work is supported by the National Key R\&D Program of China under Grant No. 2016YFA0301503 and No. 2017YFA0304500, the key NSFC grant No. 11534014 and No. 11874393.

\section{Appendix A: Calculation of winding number Eq.(\ref{WN2})}

When $E_B=0$, the definition of winding number $\nu_2$ is
\begin{equation}
\nu_2=\lim\limits_{L\rightarrow\infty}\frac{1}{4\pi i}\int_0^{2\pi}\mathrm{d}\delta\frac{1}{\partial\delta}\mathrm{ln}[\mathrm{det}(H^2)].
\label{WN2Def}
\end{equation}
Given Eqs.(\ref{ENFUN}), it can be rewritten as
\begin{eqnarray}
\nu_2
&=&\lim\limits_{L\rightarrow\infty}\frac{1}{2\pi i}\int_0^{2\pi}\mathrm{d}\delta\frac{1}{\partial\delta}\mathrm{ln}[\mathrm{det}(M+N)].
\end{eqnarray}
Above we have used the fact $(M-N)^T=(M+N)$.
When $0<\Delta/t<1$, we can define parameters
\begin{equation}
t_1e^\eta\equiv t+\Delta,t_1e^{-\eta}\equiv t-\Delta.
\end{equation}
Then the tridiagonal matrix $M+N$ turns into
\begin{equation}
M+N=\left(\begin{array}{cccc}V_1&-t_1e^\eta& &-t_1e^{-\eta}\\ -t_1e^{-\eta}&V_2&\ddots& \\  &\ddots&\ddots&-t_1e^\eta\\ -t_1e^\eta& &-t_1e^{-\eta}&V_L\end{array}\right).
\end{equation}
Now we perform a similarity transformation
\begin{equation}
S(M+N)S^{-1}=\left(\begin{array}{cccc}V_1&-t_1& &-t_1e^{-L\eta}\\ -t_1&V_2&\ddots&\\  &\ddots&\ddots&-t_1\\ -t_1e^{L\eta}& &-t_1&V_L\end{array}\right),
\end{equation}
with diagonal matrix $S=\mathrm{diag}(e^\eta,e^{2\eta},...,e^{L\eta})$.
Then in the large-$L$ limit,
\begin{equation}
\mathrm{det}(M+N)=-t_1^Le^{L\eta}+\mathrm{det}(H_1),
\end{equation}
where
\begin{equation}
H_1=\left(\begin{array}{cccc}V_1&-t_1& & \\ -t_1&V_2&\ddots& \\  &\ddots &\ddots&-t_1\\ & &-t_1&V_L\end{array}\right).
\end{equation}
Introducing a Fourier transformation
\begin{equation}
R_{n,j}=\frac{1}{\sqrt{L}}e^{2\pi i\beta nj}e^{-nh+in\delta/L},
\end{equation}
the matrix $H_1$ is transformed into
\begin{equation}
RH_1R^{-1}=\left(\begin{array}{cccc}W_1&V& &Ve^{Lh-i\delta}\\ V&W_2&\ddots & \\ \ &\ddots&\ddots&V\\ Ve^{-Lh+i\delta}& &V&W_L\end{array}\right),
\end{equation}
with $W_n=-2t_1\mathrm{cos}(2\pi \beta n)$.
Given the positive $h$, we obtain
\begin{equation}
\mathrm{det}(M+N)=-t_1^Le^{L\eta}+(-1)^{L+1}V^Le^{Lh-i\delta}+\mathrm{det}(H_2),
\end{equation}
with
\begin{equation}
H_2=\left(\begin{array}{cccc}W_1&V& & \\ V&W_2&\ddots& \\  &\ddots&\ddots&V\\ & &V&W_L\end{array}\right).
\end{equation}
The matrix $H_2$ describes the single-particle physics of the classic Hermitian AAH model under OBC, but with a parameter interchange $t_1\leftrightarrow V$.
Shown in Ref.\cite{Jiang2019}, in the large-$L$ limit
\begin{equation}
\lim\limits_{L\rightarrow\infty}|\mathrm{det}H_2|=[\mathrm{max}(t_1,V)]^L.
\end{equation}
Finally, we have
\begin{eqnarray}
\mathrm{det}(M+N)=&&-t_1^Le^{L\eta}+(-1)^{L+1}V^Le^{Lh-i\delta}\notag\\
&&+\epsilon[\mathrm{max}(t_1,V)]^L,\notag\\
=&&-t_1^Le^{L\eta}+(-1)^{L+1}V^Le^{Lh-i\delta},
\label{detH}
\end{eqnarray}
with $\epsilon$ a possible $L$-dependent sign.
Since the winding number Eq.(\ref{WN2Def}) reveals how $\mathrm{det}(H^2)$ winds around the origin in complex energy plane with respect to $\delta$ from $0$ to $2\pi$, we rewrite it with the aid of sign operator \cite{Jiang2019}
\begin{equation}
\nu_2=\frac{1}{2}\sum_i\mathrm{sgn}[x(\delta_i)]\cdot\mathrm{sgn}[\frac{dy(\delta_i)}{d\delta}],
\end{equation}
where $x=\mathrm{Re}[\mathrm{det}(M+N)]$ is the real part and $y=\mathrm{Im}[\mathrm{det}(M+N)]$ is the imaginary part.
$\delta_i$ is the $i$-th solution of $y(\delta)=0$.
From Eq.(\ref{detH}), we can obtain
\begin{eqnarray}
x&=&-(t_1e^\eta)^L+(-1)^{L+1}2V^L\mathrm{cosh}(hL)\mathrm{cos}(\delta)\notag\\
y&=&(-1)^{L}2V^L\mathrm{sinh}(hL)\mathrm{sin}(\delta).
\end{eqnarray}
In order to separate real and imaginary parts nicely, we added an infinitesimal term in $\mathrm{det}(M+N)$.
$y(\delta)=0$ has two solutions $\delta_1=0$ and $\delta_2=\pi$.
The winding number
\begin{eqnarray}
\nu_2&=&\frac{1}{2}\mathrm{sgn}[-(t_1e^\eta)^L+(-1)^{L+1}2V^L\mathrm{cosh}(hL)](-1)^{L}\notag\\
&&+\frac{1}{2}\mathrm{sgn}[-(t_1e^\eta)^L-(-1)^{L+1}2V^L\mathrm{cosh}(hL)](-1)^{L+1}\notag\\
&=&-\frac{1+\mathrm{sgn}[2V^L\mathrm{cosh}(hL)-(t_1e^\eta)^L]}{2}\notag\\
&=&-\frac{1+\mathrm{sgn}[(Ve^h)^L-(t+\Delta)^L]}{2}\notag\\
&=&-\theta(Ve^h-[t+\Delta]),
\end{eqnarray}
with $\theta$ the step function.

\section{Appendix B: Calculation of Lyapunov exponent Eq.(\ref{LEbulk})}

According to Thouless \cite{Thouless1}, the LE of an eigenstate with an energy in the neighborhood of $E_B$ is given by
\begin{equation}
\gamma=\int d\varepsilon \rho(\varepsilon)|\varepsilon-E_B|-\mathrm{ln}(t'),
\end{equation}
where $\rho(\varepsilon)$ is the density of state.
Due to the presence of p-wave pairing, we introduced a parameter $t'$ to reset the energy scale, which will be determined later.
Furthermore, numerical results show that LEs are energy-independent, and we will set $E_B=0$ for the sake of simplicity.
In order to calculate the LE, we define
\begin{eqnarray}
g&&\equiv\mathrm{ln}|\mathrm{det}(H)|=\frac{1}{2}\mathrm{ln}|\mathrm{det}(H^2)|\notag\\
&&=\mathrm{ln}|\mathrm{det}(M+N)|.
\end{eqnarray}
Indicating by $\lambda_1,...,\lambda_L$ the eigenvalues of $M+N$, we can rewrite it as
\begin{equation}
g=\sum_{n=1}^L\mathrm{ln}|\lambda_n|.
\end{equation}
In the large-$L$ limit, we replace the summation by integration and
\begin{equation}
g=L\int d\varepsilon \rho(\varepsilon)|\varepsilon|.
\end{equation}
Now we obtain
\begin{equation}
\int d\varepsilon \rho(\varepsilon)|\varepsilon|=\frac{g}{L}=\frac{\mathrm{ln}|\mathrm{det}(M+N)|}{L},
\end{equation}
and the relation between LE and determinant of Hamiltonian
\begin{equation}
\gamma=\frac{\mathrm{ln}|\mathrm{det}(M+N)|}{L}-\mathrm{ln}(t').
\end{equation}
Substituting Eq.(\ref{detH}), we have
\begin{eqnarray}
\gamma&=&\frac{\mathrm{ln}\left[|-(t+\Delta)^{L}+(-1)^{L+1}V^Le^{Lh-i\delta}|/{t'}^L\right]}{L}\notag\\
&=&\left\{\begin{array}{l}\mathrm{ln}[(t+\Delta)/t'],\quad Ve^h<t+\Delta,\\ \mathrm{ln}[Ve^h/t'],\quad\quad\quad Ve^h>t+\Delta.\end{array}\right.
\end{eqnarray}
Considering that the LE must be zero when $V=0$, we obtain $t'=t+\Delta$.
Then the LE of bulk single-particle states
\begin{equation}
\gamma=\left\{\begin{array}{l}0,\quad\quad\quad\quad\quad\quad\quad Ve^h<t+\Delta,\\ \mathrm{ln}[Ve^h/(t+\Delta)],\quad Ve^h>t+\Delta.\end{array}\right.
\end{equation}
Notice that the calculation shown in Appendixes also works when $h=0$, and the LE $\gamma$ can apply to the Hermitian case.

\end{document}